\RequirePackage{fix-cm}
\documentclass[smallextended]{svjour3}       
\smartqed  
\usepackage{graphicx}
\usepackage{epstopdf}
\usepackage{amssymb}
\usepackage{cite}
\begin{document}

\title{No Effect of Steady Rotation on Solid $^4$He in a Torsional Oscillator}


\author{M.\,J.\,Fear$^{a}$\and P.\,M.\,Walmsley$^{a}$\and D.\,E.\,Zmeev$^{a,b}$\and J.\,T.\,M\"{a}kinen$^{a,c}$
\and A.\,I.\,Golov$^{a}$}



              
\institute{	$^{a}$:School of Physics and Astronomy, The University of Manchester, Oxford Road, M13 9PL, UK.\\
			$^{b}$: Department of Physics, Lancaster University, Lancaster, LA1 4YB, UK.\\
			$^{c}$: Low Temperature Laboratory, Department of Applied Physics, Aalto University, FI-00076 AALTO, Finland\\
			}

\date{Received: date / Accepted: date}

\maketitle

\begin{abstract}
We have measured the response of a torsional oscillator containing polycrystalline hcp solid $^{4}$He to applied steady rotation in an attempt to verify the observations of several other groups that were initially interpreted as evidence for macroscopic quantum effects. The geometry of the cell was that of a simple annulus, with a fill line of relatively narrow diameter in the centre of the torsion rod. Varying the angular velocity of rotation up to 2\,rad\,s$^{-1}$ showed that  there were no step-like features in the resonant frequency or dissipation of the oscillator and no history dependence, even though we achieved the sensitivity required to detect the various effects seen in earlier experiments on other rotating cryostats. All small changes during rotation were consistent with those occurring with an empty cell. We thus observed no effects on the samples of solid $^4$He attributable to steady rotation.
\end{abstract}

\section{Introduction}

Some of the most striking properties of superfluids, such as persistent currents and quantized vortices which are due to macroscopic quantum coherence, become manifest during rotation. The responses of many different torsional oscillator (TO) experiments (undergoing AC rotation) containing solid $^4$He at temperatures below 200\,mK were thought to indicate the presence of supersolidity although it is now widely accepted that this behaviour can be explained by the temperature-dependent shear modulus of solid $^4$He (see reviews in Refs.\,\citen{Chan13,Beamish12a,Balibar12}).  The observation of further anomalous changes due to applied steady (DC) rotation in the resonant frequency and dissipation of torsional oscillators containing solid $^4$He at low temperatures was also interpreted as evidence for the existence of superflow and perhaps quantized vortices within the solid samples\cite{Gumann09,Yagi10,Choi10,Choi12,Choi12b,Takahashi12}. It was thought that superimposing DC rotation onto the oscillatory motion of a TO would allow effects due to macroscopic phase coherence (such as some form of supersolidity) to be distinguished from classical elastic effects.

The experiments utilizing DC rotation were carried out by several different research groups using two different rotating cryostats. The ISSP group\cite{Gumann09,Yagi10} who used one of their own rotating cryostats and the KAIST\cite{Choi10,Choi12,Choi12b} and Keio\cite{Takahashi12} groups both independently collaborated with the RIKEN group to use the RIKEN instrument. The initial interpretations of these experiments were all based on macroscopic quantum effects, such as some form of quantized vorticity or analogues of the de Haas-van Alphen or Shubnikov-de Haas effects. However, there are notable differences between the experiments which suggests that these observations may be due to coupling between the solid samples in the TOs and cryostat-dependent effects such as rotational noise and vibration levels. Rotating dilution refrigerators, due to their complex structure, can have mechanical resonances in either the drivetrain or supporting framework that are excited at particular values of angular velocity. The mechanical properties of solid helium mean that it is very sensitive to external perturbations (such as very low levels of vibration)\cite{Rojas10,Haziot13}. The ISSP group observed that the dissipation of their TO increased as the angular velocity, $\Omega$, was increased with no corresponding change in frequency but they also point out that their TO was not functioning reliably above 1.256\,rad\,s$^{-1}$. On the other hand, the most prominent features of experiments on the RIKEN cryostat are periodic step-like changes in the TO resonant frequency and dissipation upon sweeping the rotation velocity and also hysteresis when cycling the rotation velocity at different temperatures.

Given that these observations are still unexplained, we have used a rigid TO of relatively simple construction mounted on a recently built rotating dilution refrigerator to see if any of these phenomena could be reproduced in another laboratory. The performance of the cryostat was investigated in detail\cite{Fear13} just before the commencement of this experiment. We found that the rotation is smooth to around 1 part in $10^3$ and that the amplitude of vibration at the experimental stage below the mixing chamber is $\simeq 2$\,nm at the maximum angular velocity of 2.5\,rad\,s$^{-1}$ making this an ideal platform to use in a new search for any effect of rotation on solid $^4$He. We note that there is very little published information on the detailed performance characteristics of the other rotating cryostats used for earlier TO studies of solid $^4$He which limits our ability to make a thorough comparison of the relevant merits of the different instruments and whether they have specific features that could lead to any peculiar behavior of a TO containing solid $^4$He.

\section{Experimental setup}

\begin{figure}[h]
\begin{center}
\includegraphics[width=6cm]{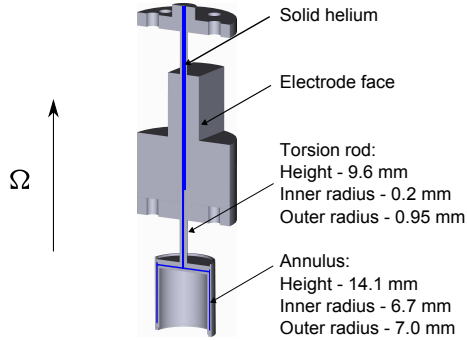}
\end{center}
\caption{(color online) Schematic of BeCu torsional oscillator used in this work. The darker shading indicates the regions occupied by solid $^{4}$He. The cell was attached to the mixing chamber of a rotating dilution refrigerator \cite{Fear13}.}
\label{fig:cellschematic}
\end{figure}

The cell consisted of a BeCu compound TO, a schematic of which is shown in Figure \ref{fig:cellschematic}. The torsion head was a simple annular geometry with thick end caps soldered in position in order to maintain the overall rigidity of the whole cell. The annulus was 14.1\,mm in height, with an inner radius of 6.7\,mm and a radial gap of 0.3\,mm. Helium was supplied to the annulus via a 0.4\,mm diameter hole centered in the 1.9\,mm diameter torsion rod. The fill line was split inside the torsion head, connecting the annulus to the pressurized line via two paths on opposite sides of the annulus. The motion of the oscillator was driven and detected capacitively using two electrodes that were positioned against a flat surface on the large isolator mass. We utilized the resonant mode where the torsion head and the large isolator mass oscillate in antiphase. This had a resonant frequency of $f_0\simeq 880$\,Hz and a $Q$ value of $\simeq 5\times10^{5}$ at low temperatures. The drive amplitude was selected such that the rim velocity of the annulus did not exceed 10\,$\mu$m\,s$^{-1}$. The moment of inertia of the larger mass was approximately 60 times larger than that of the torsion head. When the TO was mounted on the rotating cryostat, the coaxial alignment between the TO and cryostat rotation axis was better than 0.1\,mm . The cell was filled with commercial grade $^{4}$He, with a nominal natural $^{3}$He impurity of $\simeq 3\times 10^{-7}$. Starting from a temperature of 3\,K, two samples were prepared using the blocked capillary method at initial pressures of 78 and 80 bar. The inferred final pressure for these samples was approximately 47 bar\cite{Driessen86}. 

\section{Measurements at $\Omega=0$}

\begin{figure}[t]
\begin{center}
\includegraphics[width=9cm]{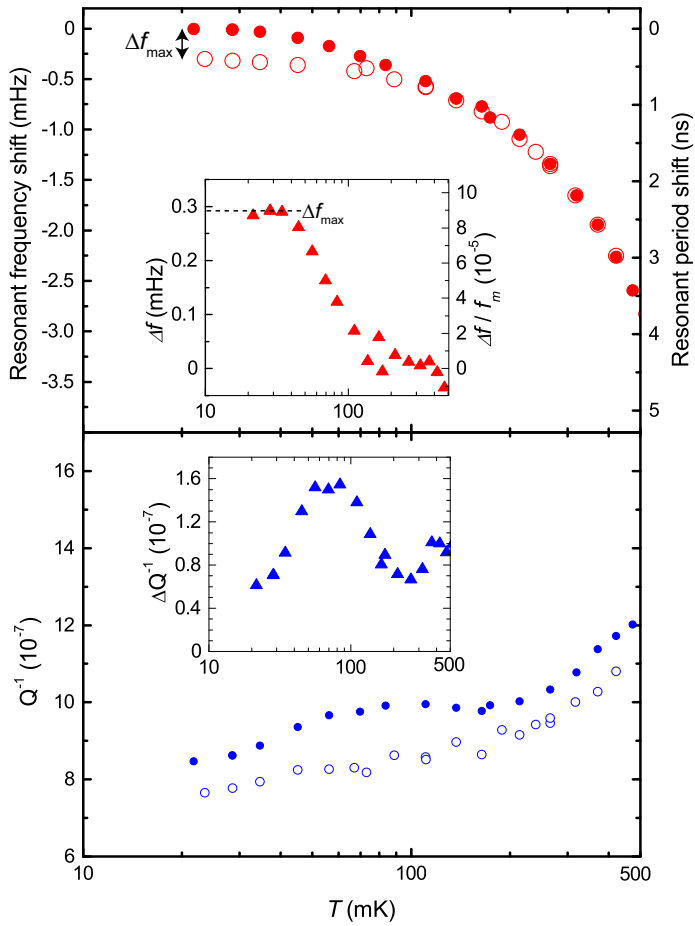}
\caption{(color online) Temperature dependence of the TO resonant frequency (top) and inverse quality factor (bottom). In both plots, open symbols show measurements for the empty cell and closed symbols are for measurements with a polycrystalline sample of solid $^4$He. The frequency shift is shown with respect to the low temperature value. The insets show the corresponding shifts due to solid helium after the empty cell data has been subtracted.}
\label{fig_temp}
\end{center}
\end{figure}

The temperature dependence of the resonant frequency and the inverse quality factor for both the empty TO and when a sample of polycrystalline hcp phase solid $^4$He was present are shown in Figure\,\ref{fig_temp}. In the case of the resonant frequency, the reduction in frequency due to the mass-loading of the solid sample, $f_m=3.25$\,Hz has been subtracted from the empty cell data. The main observation is that in the presence of solid helium, the frequency increases and the dissipation peaks at around 70\,mK, which is qualitatively similar to what was observed in many earlier TO experiments. However, the maximum normalized frequency shift, $\Delta f_\mathrm{max} / f_m \simeq 9\times 10^{-5}$, which occurs at the lowest temperatures, is very small when compared to many other TO experiments.

There are several different mechanisms through which the low temperature stiffening of solid helium can produce changes in the frequency and dissipation of TOs\cite{Beamish12,Maris12,Reppy12}. One important effect for the TO used in this work is the contribution of the helium in the torsion rod to the effective torsion constant. Beamish \textit{et al.}\cite{Beamish12} showed that this effect will increase the frequency by an amount,
\begin{equation}\label{eqn_rod}
\frac{\Delta f_{\mathrm{rod}}}{f_0}=\frac{1}{2}\frac{\mu_{\mathrm{He}}}{\mu_{\mathrm{BeCu}}}\frac{1}{(\frac{r_{o}}{r_{i}})^{4}-1},
\end{equation}
where $r_o$ and $r_i$ are the outer and inner radii of the torsion rod and $\mu_{\mathrm{He}}/\mu_{\mathrm{BeCu}}\simeq2.8\times10^{-4}$ is the ratio of the shear moduli of solid helium relative to BeCu. For our TO, the upper limit of this effect is $\Delta f_\mathrm{rod}/f_m\sim 7.4\times 10^{-5}$ which is approximately 80\% of the maximum frequency shift that we observe. It thus seems likely that most of the observed effect is due to this mechanism and that any supersolid fraction must therefore have an upper limit of $\simeq 1\times 10^{-5}$. This is consistent with recent experiments\cite{Kim14} on bulk solid helium, that used TOs with a very similar geometry to ours but with a separate fill capillary and solid torsion rod, and found an upper limit for a supersolid fraction of $4\times10^{-6}$. 

\section{Measurements during rotation}

\begin{figure}[t]
\begin{center}
\includegraphics[width=9cm]{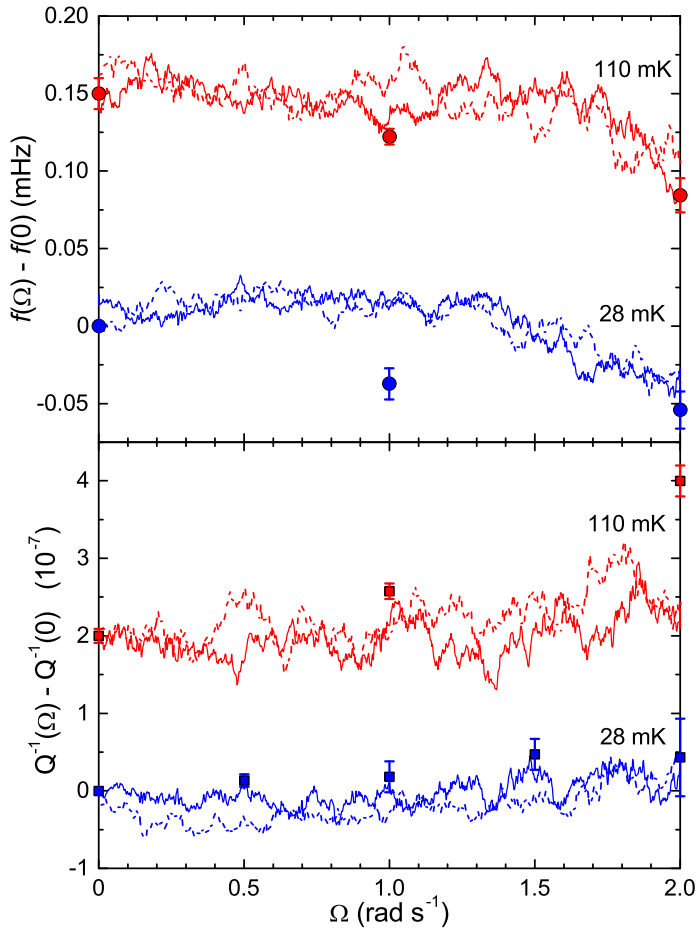}
\caption{(color online) Angular velocity dependence of the shifts in resonant frequency (top) and inverse $Q$ (bottom) relative to the stationary ($\Omega=0$) values for two different temperatures. The solid and dashed curves are for spin-down and spin-up respectively. We obtained comparable results for both of the solid $^4$He samples that were investigated. The solid symbols show the corresponding shifts for the empty TO at constant values of $\Omega$ after averaging the resonant frequency and dissipation over 10\,minutes. The error bars indicate the standard deviation. For clarity, the 110\,mK data has been shifted upwards by 0.15\,mHz in the top panel and $2\times10^{-7}$ in the bottom panel.}
\label{fig_omega}
\end{center}
\end{figure}

The main purpose of this experiment was to see if there was any effect of DC rotation on solid helium, which may indicate some form of macroscopic quantum coherence. Typical measurements of the frequency and inverse $Q$ as a function of $\Omega$ are shown in Figure\,\ref{fig_omega}. Although we tried several different measurement procedures, the data shown in Fig.\,\ref{fig_omega} were obtained by initially starting steady rotation at $\Omega=2$\,rad\,s$^{-1}$  at 500\,mK before cooling to a variable lower temperature (the examples shown in the figure are for 28 and 110\,mK) and then conducting a slow linear spin-down and subsequent spin-up (with $|\dot{\Omega}|\simeq3.4\times10^{-4}$\,rad\,s$^{-2}$). This was done to check earlier observations of hysteresis and staircase-like behavior \cite{Choi10,Choi12} when using this protocol. Cooling through a phase transition into a superfluid state while continuously rotating should give the vortex state with the lowest free energy but starting rotation while already in the superfluid state would create the vortex structures with the lowest critical velocity. In this experiment, there was no dependence on whether the sample was cooled while continuously rotating compared to starting rotation after cooling to a low temperature. Neither the frequency or dissipation showed any change compared to the stationary values for rotation with $\Omega<1.3$\,rad\,s$^{-1}$ at any temperature. The frequency was slightly lower and the dissipation larger at large values of $\Omega$ but there was no sign of any periodic staircase-like structure. There was also never any hysteresis observed when comparing spin-up to spin-down. The small shifts that we did observe at high $\Omega$ cannot be related to any $\Omega$-dependent property of solid $^{4}$He as very similar shifts are also observed when the cell is empty. It seems most likely that these empty cell changes are due to some form of coupling between rotation of the cryostat and the TO, perhaps due to the increased rotational noise and vibration levels at high $\Omega$.

We have thus not been able to reproduce any of the features observed in other TO experiments conducted on rotating cryostats. The low temperature shifts in resonant frequency and dissipation that we have found (with $\Omega=0$) are around two orders of magnitude smaller than those observed in the earlier experiments, where it seems that the various elastic effects were dominant. Even with the small shifts we have observed, our TO would have still been sensitive to the effects of DC rotation that have been reported. For example, one observation was that rotation linearly suppressed the low temperature frequency shift\cite{Choi10,Choi12} such that $\Delta f_\mathrm{max}(\Omega)=\left(1-A\Omega\right) \Delta f_\mathrm{max}(0)$ where $A\simeq0.08$\,s. However, our measurements for $\Omega\leq 1.3$\,rad\,s$^{-1}$ suggests that $A<0.02$\,s. In contrast to our work, the previous observation\cite{Choi12} of hysteresis after a rotation sweep was unlikely to be an empty cell effect. Instead, the hysteretic behaviour may have been related to the relaxation of internal stress that built up in the solid $^4$He sample, perhaps related to cryostat-specific effects such as noise on the angular velocity and coupling between mechanical resonances and the TO.

\section{Summary}

We have used a torsional oscillator that was designed to minimize the influence of the temperature-dependent elastic effects known to exist in solid helium. Very small changes in resonant frequency and dissipation that we did observe at low temperatures can be largely attributed to the stiffening of the solid inside the torsion rod. By mounting the oscillator on a rotating cryostat capable of smooth rotation with low vibration levels, we found that solid helium does not respond to DC rotation up to $\Omega=2$\,rad\,s$^{-1}$ contrary to the observations of other groups. The sensitivity of our measurements exceeded that necessary for the observation of the  effects reported previously. We have thus found no evidence of any quantum phenomena related to macroscopic phase coherence in solid $^4$He.

\begin{acknowledgements}
We thank P.  Richardson, S. Gillott and M. Sellers for their technical support and acknowledge the assistance of Louis Lu during the early stages of this project. This work was funded by the UK Engineering and Physical Sciences Research Council (EPSRC) grant numbers EP/L001446/1 and EP/I003738/1. All datasets used in this publication can be obtained from http://dx.doi.org/10.6084/m9.figshare.1613411.
\end{acknowledgements}

\end{document}